\documentclass[sigconf]{acmart}

\AtBeginDocument{%
  \providecommand\BibTeX{{%
    \normalfont B\kern-0.5em{\scshape i\kern-0.25em b}\kern-0.8em\TeX}}}

\copyrightyear{2025}
\acmYear{2025}
\setcopyright{cc}
\setcctype{by}
\acmConference[SIGIR '25]{Proceedings of the 48th International ACM SIGIR Conference on Research and Development in Information Retrieval}{July 13--18, 2025}{Padua, Italy}
\acmBooktitle{Proceedings of the 48th International ACM SIGIR Conference on Research and Development in Information Retrieval (SIGIR '25), July 13--18, 2025, Padua, Italy}
\acmDOI{10.1145/3726302.3730177}
\acmISBN{979-8-4007-1592-1/2025/07}

\usepackage{enumitem}

\title{Deep Multiple Quantization Network on Long Behavior Sequence for Click-Through Rate Prediction}

\author{Zhuoxing Wei}
\affiliation{%
  \institution{Meituan}
  \city{Beijing}
  \country{China}
}
\email{weizhuoxing@meituan.com}

\author{Qi Liu}
\affiliation{%
  \institution{University of Science and Technology of China}
  \city{Hefei}
  \country{China}
}
\email{qiliu67@mail.ustc.edu.cn}
\authornote{Corresponding author}

\author{Qingchen Xie }
\affiliation{%
  \institution{Meituan}
  \city{Beijing}
  \country{China}
}
\email{xieqingchen@meituan.com}
\renewcommand{\shortauthors}{Trovato and Tobin, et al.}

\begin{document}

\begin{abstract}

 In Click-Through Rate (CTR) prediction, the long behavior sequence, comprising the user's long period of historical interactions with items has a vital influence on assessing the user's interest in the candidate item. Existing approaches strike efficiency and effectiveness through a two-stage paradigm: first retrieving hundreds of candidate-related items and then extracting interest intensity vector through target attention. However, we argue that the discrepancy in target attention's relevance distribution between the retrieved items and the full long behavior sequence inevitably leads to a performance decline. To alleviate the discrepancy, we propose the Deep Multiple Quantization Network (DMQN) to process long behavior sequence end-to-end through compressing the long behavior sequence. Firstly, the entire spectrum of long behavior sequence will be quantized into multiple codeword sequences based on multiple independent codebooks. Hierarchical Sequential Transduction Unit is incorporated to facilitate the interaction of reduced codeword sequences. Then, attention between the candidate and multiple codeword sequences will output the interest vector. To enable online serving, intermediate representations of the codeword sequences are cached, significantly reducing latency. Our extensive experiments on both industrial and public datasets confirm the effectiveness and efficiency of DMQN. The A/B test in our advertising system shows that DMQN improves CTR by 3.5\% and RPM by 2.0\%.

\end{abstract}





\maketitle

\section{Introduction}
Click-through rate (CTR) prediction is a key stage in the industrial recommendation system (RS). Items with a higher CTR ranking will be displayed to the user. Thus, the accuracy of CTR prediction has a great influence on the evolution of RS. Existing researches have demonstrated that extracting a more accurate interest intensity vector towards the candidate from the long behavior sequence that contains long-term interactions can significantly boost the performance of CTR prediction. Current approaches~\cite{pi2019practice} typically employ a two-stage paradigm to balance efficiency and effectiveness. Approximately one hundred candidate-related items are retrieved from the long behavior sequence first. Then, target attention is performed between the candidate and retrieved items to obtain an interest vector.

The primary focus of the research lies in the first stage. Researchers propose various solutions to improve retrieved items' accuracy with lightweight modules. SIM~\cite{pi2020search} applies the item's category as a relatedness metric to retrieve items. ETA~\cite{chen2022efficient} takes hamming distance computed by efficient locality-sensitive hashing to assess and retrieve items. SDIM~\cite{cao2022sampling} uses multi-round hash collision to increase the accuracy of proxy hamming metric. TWIN~\cite{chang2023twin} solves the distribution discrepancy by make two stages share an efficient target attention network. Following research change to retrieve relevant clusters rather than original items, preserving more information about users' long-term preferences. TWIN V2~\cite{si2024twin} employs a hierarchical clustering method to group items with similar characteristics into a cluster. DGIN~\cite{liu2023deep} groups the long behavior sequence using the defined relevance key (like item\_id) to enhance efficiency.The relevance distribution gap between the full long behavior sequence and the retrieved items/clusters paradigm will inevitably lead to biased and incomplete interest estimation. 

To address retrieval-induced relevance distribution discrepancy, we propose the Deep Multiple Quantization Network (DMQN) for end-to-end long sequence processing. DMQN employs multiple learnable codebooks with codewords as cluster keys, capturing multi-aspect item features while minimizing information loss. This quantization transforms long behavior sequences into compact codeword representations. HSTU~\cite{zhai2024actionsspeaklouderwords} facilitates interactions among codeword sequences, enhancing the model's understanding of user interest structures and yielding precise preference representations. Target attention between codewords and candidates then extracts long-term interests. This end-to-end approach alleviates relevance distribution discrepancies while maintaining online efficiency. Crucially, as quantization and interaction are candidate-agnostic, intermediate representations can be precomputed and cached, enabling low-latency inference.

Overall, we make the following contributions:
\begin{itemize}[leftmargin=*]
    \item We propose the Deep Multiple Quantization Network (DMQN) to quantize the long behavior sequence into approximately a hundred learnable codewords.
    \item The HSTU is incorporated to facilitate the interaction of each short codeword sequence. It enables a more profound exploration of how different interests interact and influence each other.
    \item To evaluate the effectiveness of DMQN, we conduct extensive experiments on both industrial and public datasets confirm the effectiveness and efficiency of DMQN. The A/B test in our advertising system shows that DMQN improves CTR by 3.5\% and Revenue per Mile (RPM) by 2.0\%.
\end{itemize}

\section{Methods}

\subsection{Preliminaries}
\begin{figure*}
    \centering    
    \includegraphics[width=0.9\linewidth]{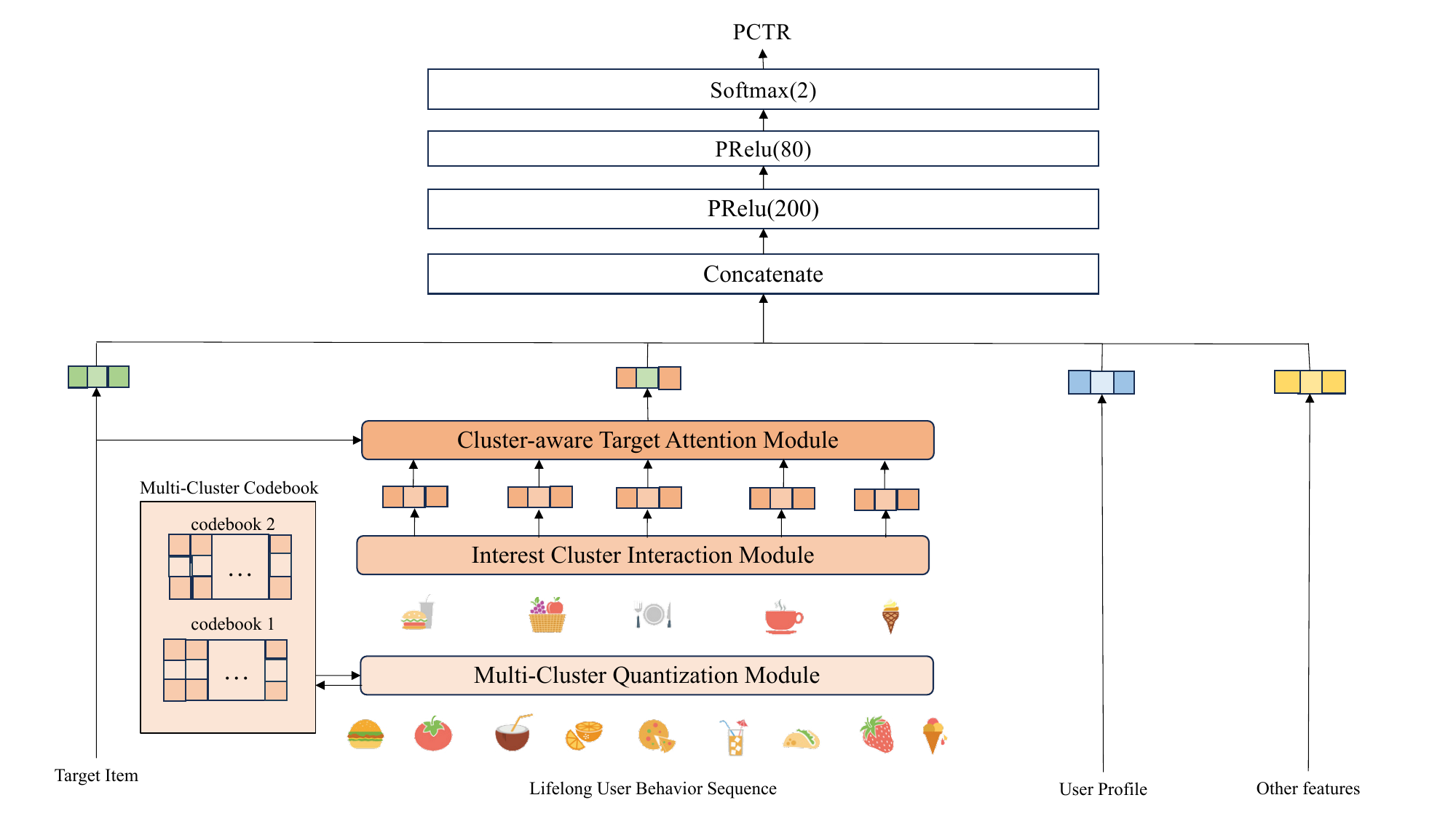}
    \caption{The overall framework of Deep Multiple Quantization Network (DMQN). DMQN consists of Multi-Cluster Quantization Module (MCQM), Interest Cluster Interaction Module(ICIM), and Cluster-aware Target Attention Module (CTAM).}
    \label{fig:DMQN}
\end{figure*}
CTR prediction focuses on estimating the likelihood that a user will click on a candidate item within a given context in the ranking stage. Each instance in this task can be represented as $(\mathbf{x}, y)$ where $\mathbf{x}=[\mathbf{x}^u, \mathbf{x}^s, \mathbf{x}^t, \mathbf{x}^c]$, $y\in\{0,1\}$ indicates click or not. $\mathbf{x}^{u}, \mathbf{x}^{s}, \mathbf{x}^t$ and $\mathbf{x}^c$ represent the features' set of user, user behavior sequence, candidate item, and context respectively. Given training dataset $D=\{(\mathbf{x}_1, y_1),...,(\mathbf{x}_N, y_N)\}$, we need to learn a model $f$ to predict the CTR, which can be formulated as the following Eq.~(\ref{eq:ctr_formulation}):
\begin{equation}
    \label{eq:ctr_formulation}
    \hat{y_i}=f(\mathbf{x}).
\end{equation}
where $\hat{y_i}$ is the estimated probability and $f$ is the CTR model.
CTR model is usually trained as a binary classification problem by minimizing the negative log-likelihood loss on the training dataset:
\begin{equation}
    \label{eq:ctr_loss}
    L_{ctr}(D)=-\frac{1}{N} \sum_{i=1}^{N}y_i \log (\hat{y_i})+(1-y_i)\log (1-\hat{y_i}).
\end{equation}
where $N$ is the size of the training dataset. For conciseness, we omit the subscript $i$ in the following description when no confusion. As shown in Figure~\ref{fig:DMQN}, DMQN is composed of the Multi-Cluster Quantization Module (MCQM), the Interest Clusters Interaction Module (ICIM), the Cluster-aware Target Attention Module (CTAM), and Multi-Layer Perception (MLP). 

\subsection{Multi-Cluster Quantization Module}
Since DMQN aims at long behavior sequence modeling, we provide detail about $\mathbf{x}^s$. The long behavior sequence consists of the user's various interactions (e.g. view, click, add-to-cart, browse-dishes, etc) with items in chronological order for a long period. There $\mathbf{x}^s$ can be represented as $\mathbf{x}^s=[\mathbf{x}^s_1, \mathbf{x}^s_2, ..., \mathbf{x}^s_L] \in R^{L \times D}$, where $L$ is the length of the long behavior sequence, $D$ is embedding dimension. The goal of this module is to transform the entire user's long behavior sequence into sequences of approximately a hundred learnable interest clusters aka codewords by executing multiple times quantization. 

\subsubsection{Multi-cluster Codebook}
Our quantization method is motivated by the idea of using codebooks to compress the embedding matrix ~\cite{chen2018learning, ge2013optimized,jegou2010product, lian2020lightrec, wu2021linear}. We encode items with a set of $C$ cluster codebooks to represent global cluster interests, each codebook contains $W$ rows codewords, where each row is a $D$-dimensional vector that serves as a cluster interest basis of the latent space. The multi-cluster codebooks can be encoded as:
\begin{equation}
    \label{eq:codebook}
    \mathbf{C} = [\mathbf{c}_1, \mathbf{c}_2, ..., \mathbf{c}_{N}]
\end{equation}
where  $\mathbf{c}_j \in R^{W \times D}$ is the j-th cluster codebook and N is the number of cluster codebooks. A linear projection is employed to map the representations of the behavior sequence into the same $D$-dimensional space as the multi-cluster codebook. This can be represented as:
\begin{equation}
    \label{eq:linear_seq}
    \mathbf{h}^s_{c_j} = \text{concat}(\mathbf{e}^s_1, \mathbf{e}^s_2, ..., \mathbf{e}^s_L) W^{c_j}_2
\end{equation}
where $\mathbf{h}^s_{c_j} \in R^{L \times D}$ is the transformed representation under the j-th cluster codebook, respectively. $W^{c_j}_2 \in R^{D \times D}$ is the corresponding weight matrix. Hereafter, the $\mathbf{c_j}$ is omitted for the sake of brevity.

\subsubsection{Multi-Cluster Quantization}
It is reasonable that multiple times quantization is performed using dot product scores instead of Euclidean distance for computational and memory efficiency. The quantization of the user's behaviors can be encoded as follows:
\begin{equation}
    \label{eq:dot_seq}
    \mathbf{score}^{s} = \text{concat}(\mathbf{h}^{s}_1, \mathbf{h}^{s}_2, ..., \mathbf{h}^{s}_L) c^T
\end{equation}
Where  $\mathbf{score}^s \in R^{L \times W}$ are the dot-product score of the user's behaviors at all rows of the cluster codebook. 

The quantized scores are typically processed using a softmax function after acquiring the scores or distances. The Gumbel-Softmax technique~\cite{jang2016categorical} is then applied to introduce Gumbel noise into the scores, enabling the probabilistic selection of cluster indices rather than argmax always choosing the highest score. This approach promotes a more uniform assignment of cluster indices. The final probability distribution can be encoded as: 
\begin{equation}
    \label{eq:softmax_seq}
    \mathbf{prob}^{s_i}_{k} = \frac{exp((log({score}^{s_i}_{k}) + g_k)/ \tau)}{\sum_{p=1}^{W}exp((log({score}^{s_i}_{p}) + g_p)/ \tau)}
\end{equation}
Where $\mathbf{prob}^{s_i}_{ k}$ are the probability of the $j$-th user's behavior at the k-th row of the cluster codebook. $g_1, ..., g_W$ are i.i.d samples drawn from Gumbel(0, 1). As the softmax temperature $\tau$ approaches 0, samples from the Gumbel-Softmax distribution become one-hot and the Gumbel-Softmax distribution converges to the categorical distribution of cluster interest.

The goal of MCQM is to obtain the cluster indices of the user's long behaviors. Following the Gumbel-Softmax trick, the argmax function is applied. This can be represented as:
\begin{equation}
    \label{eq:softmax_seq_i}
    \mathbf{z}^{s}_i = \arg\max_{k}({prob}^{s_i}_{k})
\end{equation}
\begin{equation}
    \label{eq:softmax_seq}
    \mathbf{z}^{s} = \text{concat}(\mathbf{z}^{s}_1, \mathbf{z}^{s}_2, ...,\mathbf{z}^{s}_L )
\end{equation}
Where $\mathbf{z}^{s} \in \{1,..W\}^{L}$ is the cluster indices of the  user's behaviors.

We perform pooling on the embeddings of the behavior sequence corresponding to the identical indices among them. Through this operation,  interest cluster representations are generated. Which can be encoded as:

\begin{equation}
    \label{eq:pooling_k}
    \mathbf{r}_k = \text{avgpool}({\mathbf{e}^s_i | \mathbf{z}^s_i == k, i \in \{1,..., L\} }), k \in \{1,...,W\}
\end{equation}
\begin{equation}
    \label{eq:pooling}
    \mathbf{R} = \text{concat}(\mathbf{r}_1, \mathbf{r}_2, ...,\mathbf{r}_W )
\end{equation}
Where $\mathbf{r}_k$ represents the $k$-th interest representation within the user's behavior sequence. Since a vast quantity of behaviors can be clustered and represented by a relatively small number of interest clusters, it subsequently becomes feasible to conduct a complex attention mechanism. 

\subsection{Interest Clusters Interaction Module}
In ICIM, the HSTU is incorporated to facilitate the interaction among the user's interest clusters. It enables a more profound exploration of how different interest groups interact and influence each other. This interaction mechanism enriches the model's understanding of the user's interest structure, allowing for a more accurate representation of their preferences based on historical behavior patterns. Which can be represented as:
\begin{equation}
    \label{eq:qkvu}
    U(R), V (R), Q(R), K(R) = \text{Split}(\varphi_1(f_1(R)))
\end{equation}
\begin{equation}
    \label{eq:qk}
    A(R)V(R) = \varphi_2(Q(R)K(R) + bias^p)V(R) 
\end{equation}
\begin{equation}
    \label{eq:u}
    Y(R) = f_2 (\text{Norm}(A(R)V(R)) \mathbf{\odot} U(R)) 
\end{equation}
where $f_i(X)$ denotes an MLP; we use one linear layer,
$f_i(R) = W_i(R) + b_i$ for $f_1$ and $f_2$ to reduce compute
complexity and further batches computations for queries
Q(R), keys K(R), values V(R), and gating weights U(R)
with a fused kernel; $\varphi_1$ and $\varphi_2$ denote nonlinearity, for
both of which we use SiLU; Norm is layer norm; and $bias^p$ denotes relative cluster bias.

\subsection{Cluster-aware Target Attention Module}
To extract the user's long-term interest representation, we conduct target attention between the interest cluster and the candidate item. This step is crucial as it bridges the gap between the user's existing interests, as encapsulated by the interest clusters, and the potential items that might be of interest to them.

\subsection{Caching The Intermediate Representation}
Since the output of ICIM is irrelevant to the candidate item, we can precompute the representation of each user's long behavior sequence and cache them in a key-value database such as Redis~\cite{carlson2013redis} after finishing training. When serving online, we can get the output from the key-value database based on user\_id rather than computing online which is computationally expensive. Thus, we can launch DMQN successfully to meet the strict latency requirement. 

\subsection{Complexity}
For the sake of brevity, the Multi-Cluster Quantization, which is similar to Multi-Head Attention, is omitted. The main sources of complexity are the Multi-Cluster Quantization Module and the Interest Clusters Interaction Module. The time complexity of the former is $O(L \cdot W \cdot D)$, and that of the latter is $O(N \cdot W^2 \cdot D + N \cdot W \cdot D^2)$. Therefore, the overall complexity of our method is $O(L \cdot W \cdot D + N \cdot W^2 \cdot D + N \cdot W \cdot D^2)$. Notably, it holds that $W<<L$, which indicates that the length of the user's behavior sequence far exceeds the number of interest clusters. Compared with the standard method HSTU with a complexity of $O(N \cdot L^2 \cdot D + N \cdot L \cdot D^2)$, our method has a relatively lower complexity.
\section{Experiment Setup}
\subsection{Datasets}
Experiments are performed using both industrial and publicly available datasets. The statistical characteristics of each dataset are presented in Table~\ref{tab:data_stats}. \textbf{Industry} utilized is the CTR dataset sourced from our online LBS platform. \textbf{Taobao}~\cite{zhu2019joint} is a prominent resource for CTR prediction studies, comprising user interactions from Taobao's industrial recommendation system. 
  
\begin{table}[h]
\caption{Statistics of datasets.}
\centering
\label{tab:data_stats}
\begin{tabular}{cl|c|c|c|c}
\toprule
\multicolumn{2}{c|}{Datasets}   & \multicolumn{1}{c}{\#Users} & \multicolumn{1}{c}{\#Items} & \multicolumn{1}{c}{\#Fields} & \multicolumn{1}{c}{\#Instances} \\
\midrule
\multicolumn{2}{c|}{Taobao}           & \multicolumn{1}{c}{988K}             & \multicolumn{1}{c}{4M}            & \multicolumn{1}{c}{7}               & \multicolumn{1}{c}{0.1B}  \\
\multicolumn{2}{c|}{Industry}           & \multicolumn{1}{c}{400M}             & \multicolumn{1}{c}{5M}            & \multicolumn{1}{c}{317}               & \multicolumn{1}{c}{6.6B}  \\
\bottomrule 
\end{tabular}
\end{table}

\section{Experiment Results}
\subsection{Overall Performance}

\begin{table}[tb!]
\centering
\caption{Performance of all methods on both datasets. The best result is in boldface and the second best is underlined. * indicates that the superiority to the best baseline.}
\label{tab:main_result}
\begin{tabular}{cl|c|c|c|c}
\toprule
& & \multicolumn{1}{c}{Industry} &\multicolumn{1}{c}{Taobao} \\
& & \multicolumn{1}{c}{AUC}  &\multicolumn{1}{c}{AUC}  \\
\midrule
\multicolumn{2}{c|}{DIN Small~\cite{zhou2018deep}}        & \multicolumn{1}{c}{0.7011}                           & \multicolumn{1}{c}{0.7043}             \\

\multicolumn{2}{c|}{DSIN~\cite{feng2019deep}}        & \multicolumn{1}{c}{0.7029}                              & \multicolumn{1}{c}{0.7057}               \\

\multicolumn{2}{c|}{DIEN~\cite{zhou2019deep}}       & \multicolumn{1}{c}{0.7035}                              & \multicolumn{1}{c}{0.7094}               \\

\multicolumn{2}{c|}{SIM~\cite{pi2020search}}    & \multicolumn{1}{c}{0.7043}                            & \multicolumn{1}{c}{0.7419}               \\


\multicolumn{2}{c|}{TWIN~\cite{chang2023twin}}       & \multicolumn{1}{c}{{0.7051}}      & \multicolumn{1}{c}{0.7523}      \\ 

\multicolumn{2}{c|}{SDIM~\cite{cao2022sampling}}       & \multicolumn{1}{c}{0.7073}                              & \multicolumn{1}{c}{0.7347}               \\ 

\multicolumn{2}{c|}{DIN Middle~\cite{zhou2018deep}}       & \multicolumn{1}{c}{{0.7077}}    & \multicolumn{1}{c}{0.7573}    \\ 

\multicolumn{2}{c|}{TWIN V2~\cite{si2024twin}}       & \multicolumn{1}{c}{{0.7078}}      & \multicolumn{1}{c}{0.7572}     \\ 

\multicolumn{2}{c|}{DIN Full~\cite{zhou2018deep}}       & \multicolumn{1}{c}{\underline{0.7087}}  & \multicolumn{1}{c}{\underline{0.7684}}   \\ 

\midrule

\multicolumn{2}{c|}{DMQN}       & \multicolumn{1}{c}{\textbf{0.7104}$^*$}    & \multicolumn{1}{c}{\textbf{0.7724}$^*$}   \\ 

\bottomrule 
\end{tabular}
\end{table}

Table~\ref{tab:main_result} shows the results of all methods. DMQN obtains the best performance in both the Industry and Taobao datasets, which shows the effectiveness of DMQN. There are some insightful findings from the results. 

The proposed DMQN reaches the best performance on both datasets. Compared with existing user behavior sequence modeling methods, DMQN compresses the extensive range of long user behaviors into approximately a hundred interest clusters. By incorporating the HSTU, it effectively promotes the interaction among these interest clusters. The HSTU's self-attention mechanisms enable a detailed exploration of the relationships and interdependencies between different interest groups, thereby enhancing the model's understanding of the user's interest architecture

DSIN's better performance than DIN vividly illustrates the crucial role of interaction in extracting users' session interests. As data complexity rises, DIN has limits, while DSIN's enhanced interaction dissects session interests more precisely. This underlines the need for stronger interaction. The HSTU is thus introduced. It aims to supercharge interaction via hierarchical and self-attention architectures, unearthing deeper insights and more accurate interest representations, advancing user behavior modeling.

The performance boost from the two-step long behavior sequence modeling process highlights its significance. SIM, SDIM, TWIN, and TWIN V2 outperform DIN, DIEN, and DSIN. Short sequences offer a limited view, while long ones comprehensively mirror users' complex, evolving interests and behaviors. Leveraging the full long sequence, as we plan, is crucial for deeper user understanding and application optimization.

\subsection{Ablation Study}
\begin{table}[tb!]
\centering
\caption{Results of Integrating ICIM Successively.}
\vspace{-0.4cm}
\label{tab:ablation_component}
\begin{tabular}{cl|c}
\toprule
& & \multicolumn{1}{c}{Industry} \\
& & \multicolumn{1}{c}{AUC} \\
\midrule
\multicolumn{2}{c|}{TWIN V2}                       & \multicolumn{1}{c}{0.7078}                         \\ 
\multicolumn{2}{c|}{DIN Full}                       & \multicolumn{1}{c}{0.7087}                         \\ \midrule
\multicolumn{2}{c|}{DMQN-simple}                & \multicolumn{1}{c}{0.7089}             \\ 
\multicolumn{2}{c|}{+ICIM(DMQN)}     & \multicolumn{1}{c}{\textbf{0.7103}}   \\

\bottomrule 
\end{tabular}
\end{table}
In this section, we investigate the effect of ICIM in DMQN and display the result in Table~\ref{tab:ablation_component}. The best baseline TWIN V2 and DIN Full are for comparison. In the ablation experiment where ICIM is removed from DMQN, DMQN-simple operates by compressing the entirety of the behavior sequence into roughly a hundred interest clusters and subsequently directly executing target attention. By contrast, the DMQN integrated with ICIM undertakes intricate interactions among the interest clusters before the stage of target attention. More precisely, we sequentially introduce the HSTU into the DMQN framework. The HSTU plays a crucial role in facilitating the interaction and information exchange among the interest clusters, thereby potentially augmenting the model's capacity to capture and understand the complex relationships and patterns within the user's behavior data, which could ultimately lead to improved performance and more accurate user interest representation.

\subsection{A/B Test on Performance and Cost}
We conducted an A/B test in the online LBS advertising system from 2024-05 to 2024-06 to measure the benefits of DMQN compared with the online baseline SIM Hard. DMQN was allocated \textbf{10\%} experiment serving traffic while SIM Hard held \textbf{70\%} main traffic. The online result showed the relative promotion of CTR and Revenue Per Mille (RPM) during one month's testing. DMQN achieved \textbf{3.5\%} and \textbf{2.0\%} accumulated relative promotion on the CTR and RPM respectively during the A/B test period, proving its effectiveness in the online LBS advertising system. The parameter storage costs of SIM and DMQN were \textbf{2.85 GB} and \textbf{4.00 GB} respectively. SIM had an average inference latency of \textbf{4.6 ms} and DMQN had \textbf{7.4 ms}. The resource cost bought by DMQN was negligible.

\section{Conclusion}
In this paper, we propose the DMQN for full long user behavior sequence modeling in the CTR prediction task. DMQN, consisting of a Multi-Cluster Quantization Module, an Interest Cluster Interaction Module and a Target Module, aims at extracting fine-grained comprehensive unbiased interest and psychological decision interest to achieve a deep understanding of the user's preference. To the best of our knowledge, DMQN is the first to achieve efficient end-to-end full long user behavior sequence modeling. 

\clearpage

\bibliographystyle{ACM-Reference-Format}
\bibliography{sample-base}


\begin{thebibliography}{19}


\ifx \showCODEN    \undefined \def \showCODEN     #1{\unskip}     \fi
\ifx \showDOI      \undefined \def \showDOI       #1{#1}\fi
\ifx \showISBNx    \undefined \def \showISBNx     #1{\unskip}     \fi
\ifx \showISBNxiii \undefined \def \showISBNxiii  #1{\unskip}     \fi
\ifx \showISSN     \undefined \def \showISSN      #1{\unskip}     \fi
\ifx \showLCCN     \undefined \def \showLCCN      #1{\unskip}     \fi
\ifx \shownote     \undefined \def \shownote      #1{#1}          \fi
\ifx \showarticletitle \undefined \def \showarticletitle #1{#1}   \fi
\ifx \showURL      \undefined \def \showURL       {\relax}        \fi
\providecommand\bibfield[2]{#2}
\providecommand\bibinfo[2]{#2}
\providecommand\natexlab[1]{#1}
\providecommand\showeprint[2][]{arXiv:#2}

\bibitem[Cao et~al\mbox{.}(2022)]%
        {cao2022sampling}
\bibfield{author}{\bibinfo{person}{Yue Cao}, \bibinfo{person}{Xiaojiang Zhou}, \bibinfo{person}{Jiaqi Feng}, \bibinfo{person}{Peihao Huang}, \bibinfo{person}{Yao Xiao}, \bibinfo{person}{Dayao Chen}, {and} \bibinfo{person}{Sheng Chen}.} \bibinfo{year}{2022}\natexlab{}.
\newblock \showarticletitle{Sampling is all you need on modeling long-term user behaviors for CTR prediction}. In \bibinfo{booktitle}{\emph{Proceedings of the 31st ACM International Conference on Information \& Knowledge Management}}. \bibinfo{pages}{2974--2983}.
\newblock


\bibitem[Carlson(2013)]%
        {carlson2013redis}
\bibfield{author}{\bibinfo{person}{J Carlson}.} \bibinfo{year}{2013}\natexlab{}.
\newblock \bibinfo{booktitle}{\emph{Redis in Action}}.
\newblock \bibinfo{publisher}{Manning}.
\newblock


\bibitem[Chang et~al\mbox{.}(2023)]%
        {chang2023twin}
\bibfield{author}{\bibinfo{person}{Jianxin Chang}, \bibinfo{person}{Chenbin Zhang}, \bibinfo{person}{Zhiyi Fu}, \bibinfo{person}{Xiaoxue Zang}, \bibinfo{person}{Lin Guan}, \bibinfo{person}{Jing Lu}, \bibinfo{person}{Yiqun Hui}, \bibinfo{person}{Dewei Leng}, \bibinfo{person}{Yanan Niu}, \bibinfo{person}{Yang Song}, {et~al\mbox{.}}} \bibinfo{year}{2023}\natexlab{}.
\newblock \showarticletitle{TWIN: TWo-stage Interest Network for Lifelong User Behavior Modeling in CTR Prediction at Kuaishou}.
\newblock \bibinfo{journal}{\emph{arXiv preprint arXiv:2302.02352}} (\bibinfo{year}{2023}).
\newblock


\bibitem[Chen et~al\mbox{.}(2022)]%
        {chen2022efficient}
\bibfield{author}{\bibinfo{person}{Qiwei Chen}, \bibinfo{person}{Yue Xu}, \bibinfo{person}{Changhua Pei}, \bibinfo{person}{Shanshan Lv}, \bibinfo{person}{Tao Zhuang}, {and} \bibinfo{person}{Junfeng Ge}.} \bibinfo{year}{2022}\natexlab{}.
\newblock \showarticletitle{Efficient Long Sequential User Data Modeling for Click-Through Rate Prediction}.
\newblock \bibinfo{journal}{\emph{arXiv preprint arXiv:2209.12212}} (\bibinfo{year}{2022}).
\newblock


\bibitem[Chen et~al\mbox{.}(2018)]%
        {chen2018learning}
\bibfield{author}{\bibinfo{person}{Ting Chen}, \bibinfo{person}{Martin~Renqiang Min}, {and} \bibinfo{person}{Yizhou Sun}.} \bibinfo{year}{2018}\natexlab{}.
\newblock \showarticletitle{Learning k-way d-dimensional discrete codes for compact embedding representations}. In \bibinfo{booktitle}{\emph{International Conference on Machine Learning}}. PMLR, \bibinfo{pages}{854--863}.
\newblock


\bibitem[Feng et~al\mbox{.}(2019)]%
        {feng2019deep}
\bibfield{author}{\bibinfo{person}{Yufei Feng}, \bibinfo{person}{Fuyu Lv}, \bibinfo{person}{Weichen Shen}, \bibinfo{person}{Menghan Wang}, \bibinfo{person}{Fei Sun}, \bibinfo{person}{Yu Zhu}, {and} \bibinfo{person}{Keping Yang}.} \bibinfo{year}{2019}\natexlab{}.
\newblock \showarticletitle{Deep session interest network for click-through rate prediction}.
\newblock \bibinfo{journal}{\emph{arXiv preprint arXiv:1905.06482}} (\bibinfo{year}{2019}).
\newblock


\bibitem[Ge et~al\mbox{.}(2013)]%
        {ge2013optimized}
\bibfield{author}{\bibinfo{person}{Tiezheng Ge}, \bibinfo{person}{Kaiming He}, \bibinfo{person}{Qifa Ke}, {and} \bibinfo{person}{Jian Sun}.} \bibinfo{year}{2013}\natexlab{}.
\newblock \showarticletitle{Optimized product quantization}.
\newblock \bibinfo{journal}{\emph{IEEE transactions on pattern analysis and machine intelligence}} \bibinfo{volume}{36}, \bibinfo{number}{4} (\bibinfo{year}{2013}), \bibinfo{pages}{744--755}.
\newblock


\bibitem[Jang et~al\mbox{.}(2016)]%
        {jang2016categorical}
\bibfield{author}{\bibinfo{person}{Eric Jang}, \bibinfo{person}{Shixiang Gu}, {and} \bibinfo{person}{Ben Poole}.} \bibinfo{year}{2016}\natexlab{}.
\newblock \showarticletitle{Categorical Reparameterization with Gumbel-Softmax}.
\newblock \bibinfo{journal}{\emph{arXiv preprint arXiv:1611.01144}} (\bibinfo{year}{2016}).
\newblock


\bibitem[Jegou et~al\mbox{.}(2010)]%
        {jegou2010product}
\bibfield{author}{\bibinfo{person}{Herve Jegou}, \bibinfo{person}{Matthijs Douze}, {and} \bibinfo{person}{Cordelia Schmid}.} \bibinfo{year}{2010}\natexlab{}.
\newblock \showarticletitle{Product quantization for nearest neighbor search}.
\newblock \bibinfo{journal}{\emph{IEEE transactions on pattern analysis and machine intelligence}} \bibinfo{volume}{33}, \bibinfo{number}{1} (\bibinfo{year}{2010}), \bibinfo{pages}{117--128}.
\newblock


\bibitem[Lian et~al\mbox{.}(2020)]%
        {lian2020lightrec}
\bibfield{author}{\bibinfo{person}{Defu Lian}, \bibinfo{person}{Haoyu Wang}, \bibinfo{person}{Zheng Liu}, \bibinfo{person}{Jianxun Lian}, \bibinfo{person}{Enhong Chen}, {and} \bibinfo{person}{Xing Xie}.} \bibinfo{year}{2020}\natexlab{}.
\newblock \showarticletitle{Lightrec: A memory and search-efficient recommender system}. In \bibinfo{booktitle}{\emph{Proceedings of The Web Conference 2020}}. \bibinfo{pages}{695--705}.
\newblock


\bibitem[Liu et~al\mbox{.}(2023)]%
        {liu2023deep}
\bibfield{author}{\bibinfo{person}{Qi Liu}, \bibinfo{person}{Xuyang Hou}, \bibinfo{person}{Haoran Jin}, \bibinfo{person}{Zhe Wang}, \bibinfo{person}{Defu Lian}, \bibinfo{person}{Tan Qu}, \bibinfo{person}{Jia Cheng}, \bibinfo{person}{Jun Lei}, {et~al\mbox{.}}} \bibinfo{year}{2023}\natexlab{}.
\newblock \showarticletitle{Deep Group Interest Modeling of Full Lifelong User Behaviors for CTR Prediction}.
\newblock \bibinfo{journal}{\emph{arXiv preprint arXiv:2311.10764}} (\bibinfo{year}{2023}).
\newblock


\bibitem[Pi et~al\mbox{.}(2019)]%
        {pi2019practice}
\bibfield{author}{\bibinfo{person}{Qi Pi}, \bibinfo{person}{Weijie Bian}, \bibinfo{person}{Guorui Zhou}, \bibinfo{person}{Xiaoqiang Zhu}, {and} \bibinfo{person}{Kun Gai}.} \bibinfo{year}{2019}\natexlab{}.
\newblock \showarticletitle{Practice on long sequential user behavior modeling for click-through rate prediction}. In \bibinfo{booktitle}{\emph{Proceedings of the 25th ACM SIGKDD International Conference on Knowledge Discovery \& Data Mining}}. \bibinfo{pages}{2671--2679}.
\newblock


\bibitem[Pi et~al\mbox{.}(2020)]%
        {pi2020search}
\bibfield{author}{\bibinfo{person}{Qi Pi}, \bibinfo{person}{Guorui Zhou}, \bibinfo{person}{Yujing Zhang}, \bibinfo{person}{Zhe Wang}, \bibinfo{person}{Lejian Ren}, \bibinfo{person}{Ying Fan}, \bibinfo{person}{Xiaoqiang Zhu}, {and} \bibinfo{person}{Kun Gai}.} \bibinfo{year}{2020}\natexlab{}.
\newblock \showarticletitle{Search-based user interest modeling with lifelong sequential behavior data for click-through rate prediction}. In \bibinfo{booktitle}{\emph{Proceedings of the 29th ACM International Conference on Information \& Knowledge Management}}. \bibinfo{pages}{2685--2692}.
\newblock


\bibitem[Si et~al\mbox{.}(2024)]%
        {si2024twin}
\bibfield{author}{\bibinfo{person}{Zihua Si}, \bibinfo{person}{Lin Guan}, \bibinfo{person}{ZhongXiang Sun}, \bibinfo{person}{Xiaoxue Zang}, \bibinfo{person}{Jing Lu}, \bibinfo{person}{Yiqun Hui}, \bibinfo{person}{Xingchao Cao}, \bibinfo{person}{Zeyu Yang}, \bibinfo{person}{Yichen Zheng}, \bibinfo{person}{Dewei Leng}, {et~al\mbox{.}}} \bibinfo{year}{2024}\natexlab{}.
\newblock \showarticletitle{TWIN V2: Scaling Ultra-Long User Behavior Sequence Modeling for Enhanced CTR Prediction at Kuaishou}.
\newblock \bibinfo{journal}{\emph{arXiv preprint arXiv:2407.16357}} (\bibinfo{year}{2024}).
\newblock


\bibitem[Wu et~al\mbox{.}(2021)]%
        {wu2021linear}
\bibfield{author}{\bibinfo{person}{Yongji Wu}, \bibinfo{person}{Defu Lian}, \bibinfo{person}{Neil~Zhenqiang Gong}, \bibinfo{person}{Lu Yin}, \bibinfo{person}{Mingyang Yin}, \bibinfo{person}{Jingren Zhou}, {and} \bibinfo{person}{Hongxia Yang}.} \bibinfo{year}{2021}\natexlab{}.
\newblock \showarticletitle{Linear-time self attention with codeword histogram for efficient recommendation}. In \bibinfo{booktitle}{\emph{Proceedings of the Web Conference 2021}}. \bibinfo{pages}{1262--1273}.
\newblock


\bibitem[Zhai et~al\mbox{.}(2024)]%
        {zhai2024actionsspeaklouderwords}
\bibfield{author}{\bibinfo{person}{Jiaqi Zhai}, \bibinfo{person}{Lucy Liao}, \bibinfo{person}{Xing Liu}, \bibinfo{person}{Yueming Wang}, \bibinfo{person}{Rui Li}, \bibinfo{person}{Xuan Cao}, \bibinfo{person}{Leon Gao}, \bibinfo{person}{Zhaojie Gong}, \bibinfo{person}{Fangda Gu}, \bibinfo{person}{Michael He}, \bibinfo{person}{Yinghai Lu}, {and} \bibinfo{person}{Yu Shi}.} \bibinfo{year}{2024}\natexlab{}.
\newblock \bibinfo{title}{Actions Speak Louder than Words: Trillion-Parameter Sequential Transducers for Generative Recommendations}.
\newblock
\newblock
\showeprint[arxiv]{2402.17152}~[cs.LG]
\urldef\tempurl%
\url{https://arxiv.org/abs/2402.17152}
\showURL{%
\tempurl}


\bibitem[Zhou et~al\mbox{.}(2019)]%
        {zhou2019deep}
\bibfield{author}{\bibinfo{person}{Guorui Zhou}, \bibinfo{person}{Na Mou}, \bibinfo{person}{Ying Fan}, \bibinfo{person}{Qi Pi}, \bibinfo{person}{Weijie Bian}, \bibinfo{person}{Chang Zhou}, \bibinfo{person}{Xiaoqiang Zhu}, {and} \bibinfo{person}{Kun Gai}.} \bibinfo{year}{2019}\natexlab{}.
\newblock \showarticletitle{Deep interest evolution network for click-through rate prediction}. In \bibinfo{booktitle}{\emph{Proceedings of the AAAI conference on artificial intelligence}}, Vol.~\bibinfo{volume}{33}. \bibinfo{pages}{5941--5948}.
\newblock


\bibitem[Zhou et~al\mbox{.}(2018)]%
        {zhou2018deep}
\bibfield{author}{\bibinfo{person}{Guorui Zhou}, \bibinfo{person}{Nan Mou}, \bibinfo{person}{Yukuai Fan}, \bibinfo{person}{Qiang Pi}, \bibinfo{person}{Wu Bian}, \bibinfo{person}{Xing Zhou}, {and} \bibinfo{person}{Hui Yang}.} \bibinfo{year}{2018}\natexlab{}.
\newblock \showarticletitle{Deep Interest Network for Click-Through Rate Prediction}. In \bibinfo{booktitle}{\emph{Proceedings of the 24th ACM SIGKDD International Conference on Knowledge Discovery \& Data Mining}}. \bibinfo{pages}{1059--1068}.
\newblock


\bibitem[Zhu et~al\mbox{.}(2019)]%
        {zhu2019joint}
\bibfield{author}{\bibinfo{person}{Han Zhu}, \bibinfo{person}{Daqing Chang}, \bibinfo{person}{Ziru Xu}, \bibinfo{person}{Pengye Zhang}, \bibinfo{person}{Xiang Li}, \bibinfo{person}{Jie He}, \bibinfo{person}{Han Li}, \bibinfo{person}{Jian Xu}, {and} \bibinfo{person}{Kun Gai}.} \bibinfo{year}{2019}\natexlab{}.
\newblock \showarticletitle{Joint optimization of tree-based index and deep model for recommender systems}.
\newblock \bibinfo{journal}{\emph{Advances in Neural Information Processing Systems}}  \bibinfo{volume}{32} (\bibinfo{year}{2019}).
\newblock


\end{thebibliography}

\appendix

\end{document}